\documentclass[aps,prd,onecolumn,groupedaddress,showpacs,nofootinbib,amssymb]{revtex4}
\usepackage[dvips]{graphicx}
\usepackage{amssymb}
\usepackage{amsmath}
\usepackage{graphicx,,color}
\usepackage{amsfonts}
\usepackage{bm}
\usepackage{cancel}
\usepackage{comment}
\newcommand\be{\begin{equation}}
\newcommand\ee{\end{equation}}

\allowdisplaybreaks[4]

\begin{document}

\tolerance=5000

\title{$k$-Inflation-corrected Einstein-Gauss-Bonnet Gravity with Massless Primordial Gravitons}
\author{S.D.~Odintsov,$^{1,2}$\,\thanks{odintsov@ice.cat}
V.K.~Oikonomou,$^{3,4,5}$\,\thanks{v.k.oikonomou1979@gmail.com} F.P.
Fronimos,$^{3}$\,\thanks{fotisfronimos@gmail.com}}
\affiliation{$^{1)}$ ICREA, Passeig Luis Companys, 23, 08010 Barcelona, Spain\\
$^{2)}$ Institute of Space Sciences (IEEC-CSIC) C. Can Magrans
s/n,
08193 Barcelona, Spain\\
$^{3)}$ Department of Physics, Aristotle University of
Thessaloniki, Thessaloniki 54124,
Greece\\
$^{4)}$ Laboratory for Theoretical Cosmology, Tomsk State
University of Control Systems and Radioelectronics, 634050 Tomsk,
Russia (TUSUR)\\
$^{5)}$ Tomsk State Pedagogical University, 634061 Tomsk,
Russia\\}

\tolerance=5000

\begin{abstract}
In the present paper, we study the inflationary phenomenology of a
$k$-inflation corrected Einstein-Gauss-Bonnet theory.
Non-canonical kinetic terms are known for producing Jean
instabilities or superluminal sound wave velocities in the
aforementioned era, but we demonstrate in this work that by adding
Gauss-Bonnet string corrections and assuming that the
non-canonical kinetic term $\omega X^\gamma$ is in quadratic, one
can obtain a ghost free description. Demanding compatibility with
the recent GW170817 event forces one to accept that the relation
$\ddot\xi=H\dot\xi$ for the scalar coupling function $\xi (\phi)$.
As a result, the scalar functions of the theory are revealed to be
interconnected and by assuming a specific form for one of them,
specifies immediately the other. Here, we shall assume that the
scalar potential is directly derivable from the equations of
motion, once the Gauss-Bonnet coupling is appropriately chosen,
but obviously the opposite is feasible as well. As a result, each
term entering the equations of motion, can be written in terms of
the scalar field and a relatively tractable phenomenology is
produced. For quadratic kinetic terms, the resulting scalar
potential is quite elegant functionally. Different exponents,
which lead to either a more perplexed solution for the scalar
potential, are still a possibility which was not further studied.
We also discuss in brief the non-Gaussianities issue under the
slow-roll and constant-roll conditions holding true, and we
demonstrate that the predicted amount of non-Gaussianities is
significantly enhanced in comparison to the $k$-inflation free
Einstein-Gauss-Bonnet theory.
\end{abstract}

%PACS numbers: 04.50.Kd, 95.36.+x, 98.80.-k, 98.80.Cq

\maketitle

\section{Introduction}

Describing the inflationary era in a consistent way is one of the
main task in modern theoretical cosmology. To date, there are two
main ways to describe the inflationary era, with the first being
by using the single scalar field description
\cite{Guth:1980zm,Linde:1993cn,Linde:1983gd}, while the other is
by using the modified gravity description
\cite{Nojiri:2017ncd,Nojiri:2010wj,Nojiri:2006ri,Capozziello:2011et,Capozziello:2010zz,Olmo:2011uz}.
Both ways are appealing, however the modified gravity description
is somewhat more complete since it offers the appealing
possibility of describing the inflationary and the dark energy era
by using the same theoretical framework
\cite{Nojiri:2003ft,Odintsov:2020iui,Odintsov:2020nwm,Sa:2020qfd}.
Of course we need also to mention that bouncing cosmology is an
also quite appealing alternative to inflation \cite{bounces}. On
the other hand, string theory is to date the most complete high
energy completion of general relativity and of the Standard Model,
and it is therefore natural to assume that it will have some
imprints on the low-energy inflationary Lagrangian. It is a well
known fact that the inflationary era is essentially a classical
theory, in which era, the Universe is four dimensional and also
evolves in a classical way, with quantum effects affecting the
evolution in a subdominant way. In this sense,
Einstein-Gauss-Bonnet theories
\cite{Hwang:2005hb,Nojiri:2006je,Cognola:2006sp,Nojiri:2005vv,Nojiri:2005jg,Satoh:2007gn,Bamba:2014zoa,Yi:2018gse,Guo:2009uk,Guo:2010jr,Jiang:2013gza,Kanti:2015pda,vandeBruck:2017voa,Kanti:1998jd,Pozdeeva:2020apf,Fomin:2020hfh,DeLaurentis:2015fea,Chervon:2019sey,Nozari:2017rta,Odintsov:2018zhw,Kawai:1998ab,Yi:2018dhl,vandeBruck:2016xvt,Kleihaus:2019rbg,Bakopoulos:2019tvc,Maeda:2011zn,Bakopoulos:2020dfg,Ai:2020peo,Odintsov:2019clh,Oikonomou:2020oil,Odintsov:2020xji,Oikonomou:2020sij,Odintsov:2020zkl,Odintsov:2020sqy,Odintsov:2020mkz,Easther:1996yd,Antoniadis:1993jc,Antoniadis:1990uu,Kanti:1995vq,Kanti:1997br}.
have an elevated role in the description of the inflationary era,
since these theories provide a string-motivated modification of
the canonical scalar field inflationary theory, with the
low-energy string corrections appearing as Gauss-Bonnet
corrections. However, the GW170817 event back in 2017
\cite{GBM:2017lvd} altered the perception of the viability of a
modified gravity theory, since, the gravitational wave speed was
found to have the same propagation speed as that of lights. Since
there is no fundamental reason coming from particle physics which
indicates that the graviton should change its mass, at least to
our knowledge, it is natural to assume that the primordial
gravitational waves should have a propagation speed equal to unity
in natural units. This sole constraint has already excluded many
theories that predicted a primordial tensor mode spectrum with
propagation speed different that that of light's, see for example
\cite{Ezquiaga:2017ekz}, and Einstein-Gauss-Bonnet theories belong
to this class of excluded theories. In some recent works we
introduced a new formalism that offers a remedy to the
non-viability issue of Einstein-Gauss-Bonnet theories and related
theories, in view of the GW170817 event
\cite{Odintsov:2019clh,Oikonomou:2020oil,Odintsov:2020xji,Odintsov:2020zkl,Odintsov:2020sqy,Oikonomou:2020sij,Odintsov:2020mkz}.
The basic new feature of these new theories which are GW170817
compatible, is that the scalar potential and the Gauss-Bonnet
scalar coupling function should no longer be considered as
independent functions that can be freely chosen, but these must be
related in a specific way.

In this work, we extend the theoretical framework of
GW170817-compatible Einstein-Gauss-Bonnet theories, to include
$k$-inflation corrections
\cite{ArmendarizPicon:1999rj,Chiba:1999ka,ArmendarizPicon:2000dh,Matsumoto:2010uv,ArmendarizPicon:2000ah,Chiba:2002mw,Malquarti:2003nn,Malquarti:2003hn,Chimento:2003zf,Chimento:2003ta,Scherrer:2004au,Aguirregabiria:2004te,ArmendarizPicon:2005nz,Abramo:2005be,Rendall:2005fv,Bruneton:2006gf,dePutter:2007ny,Babichev:2007dw,Deffayet:2011gz,Kan:2018odq,Unnikrishnan:2012zu,Li:2012vta},
thus non-canonical higher order kinetic terms for the scalar
field. The motivation for this addition is three fold: Firstly in
order to see whether a viable inflationary phenomenology can be
generated in this case, secondly in order to see whether the
non-Gaussianities are enhanced in this case, and thirdly,
$k$-inflation might offer the possibility of also describing the
dark energy era. The focus in this paper will be on the first two
aforementioned issues, namely the inflationary phenomenology and
the non-Gaussianities issue, and as we demonstrate, it is possible
to produce viable inflationary phenomenology, and also the
non-Gaussianities of the primordial power spectrum are
significantly enhanced in comparison to the $k$-inflation free
Einstein-Gauss-Bonnet theory.

\section{Einstein-Gauss-Bonnet $k$-Inflation Dynamics}

We commence by defining the gravitational action corresponding to
an $k$-Inflation-corrected Einstein-Gauss-Bonnet gravity, which
is,
\begin{equation}
\centering
\label{action}
S=\int{d^4x\sqrt{-g}\left(\frac{R}{2\kappa^2}-\alpha X-\omega X^\gamma-V(\phi)-\xi(\phi)\mathcal{G}\right)},\,
\end{equation}
with $R$ being the Ricci scalar, $\kappa=\frac{1}{M_P}$ the
gravitational constant and $M_P$ the reduced Planck mass, while
$X=\frac{1}{2}g^{\mu\nu}\partial_\mu\phi\partial_\nu\phi$ and
$V(\phi)$ are the kinetic term and scalar potential. Also
$\mathcal{G}$ denotes the Gauss-Bonnet invariant defined as
$\mathcal{G}=R^2-4R_{\mu\nu}R^{\mu\nu}+R_{\mu\nu\sigma\rho}R^{\mu\nu\sigma\rho}$
with $R_{\mu\nu}$ and $R_{\mu\nu\sigma\rho}$ being the Ricci and
Riemann tensor respectively and lastly, $\xi(\phi)$ signifies the
Gauss-Bonnet scalar coupling function. Concerning the
non-canonical kinetic term, we mention that $\omega$ is a constant
with mass dimensions [m]$^{4-4\gamma}$ for consistency. Similarly,
$\alpha$ shall take the values $ 0, 1$ but we shall leave it as it
is in order to have the phantom case available for the reader. In
the present paper, we shall also assume that the cosmological
background corresponds to that of a flat Friedman-Robertson-Walker
(FRW) metric, with the line element being,
\begin{equation}
\centering
\label{metric}
ds^2=-dt^2+a^2(t)\delta_{ij}dx^idx^j,\,
\end{equation}
where $a(t)$ denotes the scale factor. In consequence, the Ricci
and Gauss-Bonnet scalars are written as $R=12H^2+6\dot H$ and
$\mathcal{G}=24H^2(\dot H+H^2)$, with $H$ being Hubble's parameter
$H=\frac{\dot a}{a}$ and the ``dot'' signifies differentiation
with respect to the cosmic time. Furthermore, by assuming that the
scalar field is homogeneous, then the kinetic term is simplified
greatly as now $X=-\frac{1}{2}\dot\phi^2$. Due to this definition,
we shall limit our work to only integer values for $\gamma$ in
order to avoid the emergence of complex numbers.

Theories with non-canonical kinetic terms are known for producing
a formula for the velocity of the gravitational waves with does
not necessarily coincide with the speed of light. Before we
proceed with the equations of motion, it is worth taking care of
the constraints imposed by the GW170817 event. In order for this
particular model not to be at variance with the GW170817 event, we
demand that in natural units where $c=1$, the velocity of the
gravitational waves is equal to unity. Thus,
\begin{equation}
\centering
\label{cT}
c_T^2=1-\frac{Q_f}{2Q_t},\,
\end{equation}
must be equated to unity. Here, we make use of certain auxiliary
functions defined as $Q_f=16(\ddot\xi-H\dot\xi)$ and
$Q_t=\frac{1}{\kappa^2}-8\dot\xi H$. As a result, compatibility is
restored once the numerator of the second term is equated to zero,
meaning that $\ddot\xi=H\dot\xi$. This case was studied in
\cite{Odintsov:2020sqy} for the slow-roll case and in turn, the
following differential equation was found to constraint the
functional form of the Gauss-Bonnet coupling,
\begin{equation}
\centering
\label{dotphi}
\dot\phi\simeq H\frac{\xi'}{\xi''}\, .
\end{equation}
where ``prime'' denotes differentiation with respect to the scalar
field $\phi$.

The gravitational action is a powerful tool since it contains
stored all the available information about the inflationary era.
Implementing the variation principle with respect to the metric
tensor and the scalar potential, produces the gravitational field
equation and also the continuity equation of the scalar field. By
separating the equations of motion in time and space components,
the equations of motion are derived which read,
\begin{equation}
\centering
\label{motion1}
\frac{3H^2}{\kappa^2}=-\alpha X+\omega(1-2\gamma)X^\gamma+V(\phi)+24\dot\xi H^3,\,
\end{equation}
\begin{equation}
\centering
\label{motion2}
\dot H=\kappa^2(\alpha X+\omega\gamma X^\gamma+8\dot\xi H\dot H),\,
\end{equation}
\begin{equation}
\centering
\label{motion3}
\alpha(\ddot\phi+3H\dot\phi)+(\ddot\phi(2\gamma-1)+3H\dot\phi)\omega\gamma X^{\gamma-1}+V'+\xi'\mathcal{G}=0,\,
\end{equation}
where for $\gamma\to1$ one obtains the usual Einstein-Gauss-Bonnet
equations of motion. This is the set of differential equations
that must be solved in order to extract results for a given model
during the inflationary era. However, this model is hard to solve
analytically, hence, certain approximations must be made in order
to proceed. Here, we shall make two kinds of approximations. The
first is the slow-roll approximation, where one assumes that the
scalar field has an inferior canonical kinetic term compared to
the scalar potential. This approximation will be extended to the
non-canonical kinetic term as well, hence the slow-roll
approximations are,
\begin{align}
\centering
\label{slow-roll}
\dot H&\ll H^2, &\alpha X+\omega X^\gamma&\ll V&\ddot\phi&\ll H\dot\phi.\,
\end{align}
These inequalities refer to the order of magnitude and not the
sign of each term. Subsequently, the equations of motion can be
simplified greatly,
\begin{equation}
\centering
\label{motion4}
H^2\simeq\frac{\kappa^2V}{3},\,
\end{equation}
\begin{equation}
\centering
\label{motion5}
\dot H\simeq\kappa^2(\alpha X+\omega\gamma X^\gamma).\,
\end{equation}
\begin{equation}
\centering
\label{motion6}
V'+3H\dot\phi(\alpha+\omega\gamma X^{\gamma-1})\simeq0,\,
\end{equation}
Let us now proceed with the designation of certain auxiliary
parameters which shall play a significant role in subsequent
calculations. These parameters are slowly varying obviously.
Firstly, we define the set of the slow-roll indices as,
\begin{align}
\centering
\label{indices}
\epsilon_1&=-\frac{\dot H}{H^2}, &\epsilon_2&=\frac{\ddot\phi}{H\dot\phi},&\epsilon_3&=0,&\epsilon_5&=\frac{Q_a}{2HQ_t},&\epsilon_6&=\frac{\dot Q_t}{2HQ_t},\,
\end{align}
where  $Q_a=-8\dot\xi H^2$ and again,
$Q_t=\frac{1}{\kappa^2}-8\dot\xi H$. Here, the index $\epsilon_4$
was omitted simply because it is of no interest. The same can be
said about indices $\epsilon_5$ and $\epsilon_6$ but since they
are derived from the string corrections, they are presented for
the sake of completeness. Moreover, the functions which shall be
utilized have the following form,
\begin{align}
\centering
\delta_\xi&=\kappa^2 H\dot\xi,&\epsilon_s&=\epsilon _1-4\delta_\xi,&\eta_s&=\frac{\dot\epsilon_s}{H\epsilon_s},&s&=\frac{\dot c_s}{Hc_s}.
\end{align}
These functions are of paramount importance since they are
strongly connected with the observed quantities as we shall show
in the following. Before we proceed with the observational
quantities however, we note that $c_s$ is the field propagation
velocity defined as,
\begin{equation}
\centering
\label{soundwave}
c_s^2=\kappa^4\frac{2\omega_1^2\omega_2H-\omega_2^2\omega_1+4\omega_1\dot\omega_1\omega_2-2\omega_1^2\dot\omega_2}{4\Sigma}
\end{equation}
where,
\begin{equation}
\centering
\Sigma=\kappa^4\frac{\omega_1(4\omega_1\omega_3+9\omega_2^2)}{12},\,
\end{equation}
\begin{equation}
\centering
\omega_1=Q_t,\,
\end{equation}
\begin{equation}
\centering
\omega_2=\frac{2H}{\kappa^2}-24H^2\dot\xi,\,
\end{equation}
\begin{equation}
\centering
\omega_3=-\left(\frac{3H}{\kappa}\right)^2-3\alpha X-3\omega\gamma(2\gamma-1)X^\gamma+144H^3\dot\xi,\,
\end{equation}
Finally, the predicted amount of non-Gaussianities for each model
can be evaluated from the equilateral non-linear term
\cite{DeFelice:2011zh},
\begin{equation}
\centering
\label{fNL}
f_{NL}^{eq}=\frac{85}{324}\left(1-\frac{1}{c_s^2}\right)-\frac{10}{81}\frac{\lambda}{\Sigma}+\frac{55}{36}\frac{\epsilon_s}{c_s^2}+\frac{5}{12}\frac{\eta_s}{c_s^2}-\frac{85}{54}\frac{s}{c_s^2}-\frac{10}{81}\delta_\xi\left(2-\frac{29}{c_s^2}\right),\,
\end{equation}
with $\lambda$ being,
\begin{equation}
\centering
\lambda=-\omega\gamma(\gamma-1)X^{\gamma}\left(1+\frac{2(\gamma-2)}{3}\right).\,
\end{equation}
In the case of Einstein-Gauss-Bonnet gravity with canonical
kinetic term, the expected deviation from the Gaussian
distribution is small while on the other hand, the
non-Gaussianities are now expected to be enhanced due to the
inclusion of a non-canonical kinetic term. Finally, the observed
quantities which shall be evaluated and compared to the recent
Planck 2018 data are the scalar spectral index of primordial
curvature perturbations $n_S$, the spectral index of tensor
perturbations $n_T$ and the tensor-to-scalar ratio $r$, connected
to the previously defined slow-roll parameters as shown below,
\begin{align}
\centering
n_S&=1-2\epsilon_1-\eta_s-s,&n_T&=-2\epsilon_1,&r&=16\epsilon_sc_s,\,
\end{align}
with the corresponding values being
\begin{align}
\centering
n_S&=0.9649\pm0.0042,  68\% C.L&r&<0.064,  95\% C.L,\,
\end{align}
The main goal is to evaluate those quantities during the first
horizon crossing by finding the initial value of the scalar field.
One can evaluate the final value of the scalar field in the first
place by simply assuming that the slow-roll index $\epsilon_1$
becomes of order $\mathcal{O}(1)$. The initial value in turn can
be evaluated from the $e$-foldings number, since in this
framework, it is equal to,
\begin{equation}
\centering
\label{efolds}
N=\int_{\phi_i}^{\phi_f}{\frac{\xi''}{\xi'}d\phi}\, ,
\end{equation}
Although the procedure seems tedious, one must be wary not to
produce Jeans instabilities, meaning field propagation velocities
which obey the relation $c_s^2<0$ and also superluminal velocities
where $c_s>1$. The latter case does not respect causality, but
there exist certain cases which produce such result, therefore
they should be avoided. In the following, we shall study simple
coupling functions which manage to simplify the ratio $\xi'/\xi''$
which appears in the equations of motion. These functions were
proved to be capable of producing a viable phenomenology in the
canonical case.

\section{Confronting String Corrected k-Inflation with Observations}

In this section, we present a variety of models which have been
studied in previous works of ours
\cite{Odintsov:2019clh,Oikonomou:2020oil,Odintsov:2020xji,Odintsov:2020zkl,Odintsov:2020sqy,Oikonomou:2020sij,Odintsov:2020mkz},
using the canonical kinetic term in Einstein-Gauss-Bonnet gravity
theory. We shall also show that apart from being viable, the
models are also capable of predicting a larger amount of
non-Gaussianities which is an important distinction between the
$k$-inflation and the ordinary Einstein-Gauss-Bonnet models.

\subsection{ Error Function Model in the Presence of a Higher Order Kinetic Term}

Let us assume that the coupling function is that of Eq.
(\ref{xiB}) with $\alpha=1$ and $\gamma=2$. Also, in order to
avoid confusion between the models, we shall use subscripts in
similar parameters. The subscript in the first model is 1. Hence,
let the Gauss-Bonnet coupling be,
\begin{equation}
\centering
\label{xiA}
\xi(\phi)=\frac{2\Lambda_1}{\sqrt{\pi}}\int_{0}^{\delta_1\kappa\phi}{e^{-x^2}dx}\, ,
\end{equation}
As a result, Eq. (\ref{motion6}) produces the following scalar
potential,
\begin{equation}
\centering
\label{VA}
V(\phi)=\frac{12\delta_1^4(\alpha-4\delta_1^2)}{\omega}(\kappa\phi)^2\, ,
\end{equation}
where the integration constant is set equal to zero for
simplicity. In turns out that the scalar potential has an elegant
power-law form and specifically it follows a squared law form.
Then consequently,
\begin{equation}
\centering
\label{index1C}
\epsilon_1=\frac{1}{2(\delta_3\kappa\phi)^2}\, ,
\end{equation}
\begin{equation}
\centering
\label{index2A}
\epsilon_2=0\, ,
\end{equation}
\begin{equation}
\centering
\label{index5A}
\epsilon_5=\frac{4 \Lambda_1 \kappa^4 V(\phi )}{3 \sqrt{\pi } \delta_1 \kappa \phi  e^{(\delta_1 \kappa \phi)^2}+8 \Lambda_1 \kappa^4 V(\phi )}\, ,
\end{equation}
\begin{equation}
\centering
\label{index6A}
\epsilon_6=\frac{2 \Lambda_1  \left(\kappa^4V(\phi)(2 (\delta_1 \kappa \phi)^2+1)-\kappa\phi \kappa^3 V'(\phi )\right)}{(\delta_1 \kappa \phi)^2 \left(3 \sqrt{\pi } \delta_1 \kappa \phi  e^{(\delta_1 \kappa \phi)^2}+8  \Lambda_1 \kappa^4 V(\phi )\right)}\, ,
\end{equation}
\begin{equation}
\centering
\label{dxiA}
\delta_\xi=-\frac{\kappa^3\xi'(\phi)V(\phi)}{6\delta_1^2\kappa\phi}\, ,
\end{equation}
\begin{equation}
\centering
\label{lambdaA}
\lambda=-\frac{\omega}{2}\left(\frac{V(\phi)}{12(\delta_3^2\kappa\phi)^2}\right)^2\, ,
\end{equation}
These are the the only simple auxiliary parameters of the model
and each quantity can be constructed from them. It is worth
mentioning that $\epsilon_2=0$ or in other words $\ddot\phi=0$.
Thus, $\phi(t)$ is a simple linear function of time and hence the
time duration of the inflationary era can be extracted for the
initial and final value of the scalar field. These values can be
evaluated from the condition $\epsilon_1(\phi_f)=1$, which
terminates inflation and afterwards the value $\phi_i$ if the
scalar field, during the first horizon crossing, can be derived
from the $e$-folding number (\ref{efolds}). As a result, we obtain
the following expressions,
\begin{equation}
\centering
\label{scalarfA}
\phi_f=\pm\frac{1}{\sqrt{2}\delta_1\kappa}\, ,
\end{equation}
\begin{equation}
\centering
\label{scalariA}
\phi_i=\pm\frac{\sqrt{N+(\delta_1\kappa\phi_f)^2}}{\delta_1\kappa}\, .
\end{equation}
We shall use only the positive values. Assigning the values
($\omega$, $\Lambda_1$, $N$, $\delta_1$)=(-0.001, 1, 60, 0.1), in
reduced Planck Units, then $n_S=0.966942$, $n_T=-0.0165289$,
$r=0.026526$, $f_{NL}^{eq}=-5.6528$ and $c_s=0.200603$, which are
compatible results with the recent Planck 2018 data
\cite{Akrami:2018odb} and also the model is free from ghost
degrees of freedom, for these values. In addition,
$\phi_i=77.7817$, $\phi_f=7.07107$ and moreover,
$\epsilon_1=0.00826$, $\epsilon_2=4.4\cdot10^{-16}$,
$\epsilon_5=-3.5\cdot10^{-24}=\epsilon_6$ which indicates that the
slow-roll approximations hold true. Lastly,
$\delta_\xi=8.95\cdot10^{-25}$, $\epsilon_s=\epsilon_1$,
$s=3.39\cdot10^{-17}$, $\eta_s=0.0165289$. When it comes to the
validity of the necessary approximations, we note that $\dot
H/H^2\sim\mathcal{O}(10^{-3})$ as mentioned before, $\alpha
X+\omega X^\gamma\sim\mathcal{O}(10)$ and
$\ddot\phi/H\dot\phi\sim\mathcal{O}(10^{-16})$, which is
numerically expected as $\epsilon_2=0$. Furthermore, the string
corrections can be neglected since $24\dot\xi
H^3\sim\mathcal{O}(10^{-20})$, $16\dot\xi H\dot
H\sim\mathcal{O}(10^{-22})$ and $24\xi'
H^4\sim\mathcal{O}(10^{-20})$ while $V'\sim\mathcal{O}(10^2)$, so
all the approximations are justifiable. Our results are also
supported by Figs. 1 and 2 for some sets of values of the free
parameters.
\begin{figure}[h!]
\centering \label{plot1}
\includegraphics[width=20pc]{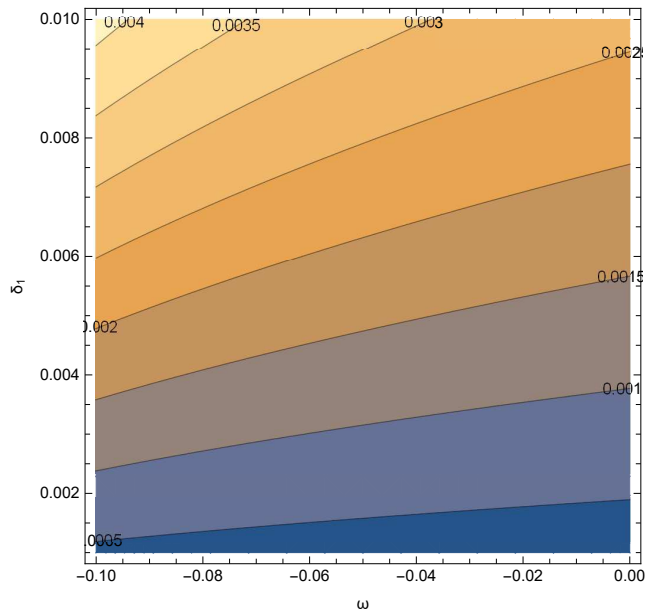}
\caption{Tensor-to-scalar ratio $r$ depending on the parameters $\omega$ and exponent $\delta_1$ ranging from $[-0.01,-0.0001]$ and [0.001,0.01] respectively.}
\end{figure}
\begin{figure}[h!]
\centering
\label{plot2}
\includegraphics[width=20pc]{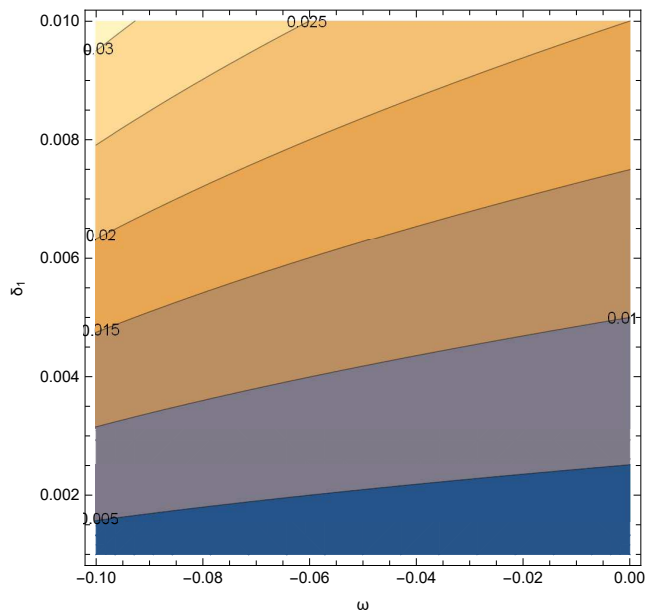}
\caption{ Field propagation velocity $c_s$ during the first
horizon crossing depending on parameters $\omega$ and $\delta_1$
ranging from $[-0.01,-0.0001]$ and [0.001,0.01] respectively. It can
easily be inferred that these values respect causality.}
\end{figure}
The model under study unfortunately should be rendered as
intrinsically unrealistic. This is because $c_s=1.31198$ during
the final stage of the inflationary era, which was not the case
with the models studied in our previous works. In consequence,
causality is violated and hence it is apparent that for some
values of the free parameters, ghost degrees of freedom make their
presence apparent. However, slight change in the numerical value
of the non canonical kinetic term restores viability. This time,
for $\omega=10^{-8}$ we have $n_S=0.966942$, $r=0.0264463$,
$n_T=-0.0165289$ and $f_{NL}^{eq}=-5.68995$ while $c_s=0.2$ and
$c_s=0.999998$ during the initial and final stage of the
inflationary era respectively. In this case, causality is restored
while the approximations assumed previously still apply. This is
indicative of the fact that the free parameters of the theory must
be chosen properly in certain cases in order to achieve
compatibility. One should not naively designate the free parameters
for even a slight detail as the one present in this paper, the sound
speed at the end of the inflationary era, can turn the model into
unrealistic. The behavior of the speed $c_s$ depending on
various parameters can be found in Fig. 2. The only difference
is the scalar potential. The same applies to the constant-roll
case which shall be studied in the following section.

As a last comment, it would be wise to address the issue of the
free parameters, and especially their range of values and the
overall impact on the approximations made. In this first model,
the only parameters that affect the results significantly are the
ones showcased in the previous diagrams, namely the non
canonical parameter $\omega$ and the exponent $\delta_1$
of the error function Gauss-Bonnet coupling. For the non
canonical parameter $\omega$, it is essential to assume that
its value is positive in this case in order to achieve a non superluminal
velocity at the end of the inflationary era only, otherwise everything
is in aggrement with the Planck 2018 data, whereas a negative value could still
result in a viable $c_s$ only if $\delta$ is increased, however this is prohibited
 since $r$ becomes greater than 0.064 and thus cannot be accepted.
 A decisive factor as mentioned before is exponent $\delta_1$ which
 for greater values,for instance $\delta_1=-0.3$, the tensor-to-scalar ratio becomes
equal to $r=0.07933$ which is obviously not an acceptable value.
Moreover, such an increase results in a decrease in the equilateral
nonlinear term to $f_{NL}^{eq}=-0.3332$. On the other hand, a
decrease of the aforementioned exponent is acceptable since it
decreases the tensor-to-scalar ratio and results in a rather significant
increase on the amount of non Gaussianities. As an example, we mention
that $\delta_1=-0.03$ generates an $f_{NL}^{eq}=-67.0681$ while the
sound wave velocity during the initial and final stage remains real and below
unity. Finally, the order of the Gauss-Bonnet coupling, namely the free
parameter $\lambda_1$ is not so significant. It turns out that no observed
index is dependent on it however it is worth mentioning that the sound
speed at the final change can be altered from $\lambda_1$, only if it obtains
a quite large value, for instance $\lambda_1\sim\mathcal{O}(10^{12})$ and
beyond. Note however that an extreme value of $\mathcal{O}(10^{24})$
and greater would result in a direct violation of the approximations.

\section{Einstein-Gauss-Bonnet $k$-Inflation Phenomenology with Constant-Roll Conditions}

In the previous examples we showcased how the slow-roll
assumptions facilitate our study greatly. In general, the fact
that $\ddot\phi\ll H\dot\phi$ simplifies the continuity equation
and an elegant analytic form for the scalar potential can be
extracted. Another condition which has the same effect is the
constant-roll evolution for the scalar field. This is a quite
interesting case since in its core is a theory which can
potentially produce non-Gaussianities in the primordial power
spectrum, thus an enhanced equilateral non-linear term
$f_{NL}^{eq}$ is naturally produced. Let us commence our study by
introducing the constant-roll condition, that is $\ddot\phi=\beta
H\dot\phi$ where $\beta$ is the constant-roll parameter.
Essentially, from the previous results, it becomes apparent that
$\epsilon_2=\beta$ hence we shall limit our work to only small
values of $\beta$. From Eq. (\ref{cT}), one obtains the new time
derivative for the scalar field, which is written as,
\begin{equation}
\centering
\label{dotphi2}
\dot\phi=H\frac{(1-\beta)\xi'}{\xi''}\, ,
\end{equation}
which as expected has the same limit as Eq. (\ref{dotphi}) for
$\beta\to0$. Consequently, the equations of motion are altered as
follows,
\begin{equation}
\centering
\label{motion7}
H^2\simeq\frac{\kappa^2V}{3},\,
\end{equation}
\begin{equation}
\centering
\label{motion8}
\dot H\simeq\kappa^2(\alpha X+\omega\gamma X^\gamma).\,
\end{equation}
\begin{equation}
\centering
\label{motion9}
V'+H\dot\phi(\alpha(3+\beta)+\omega\gamma(\beta(2\gamma-1)+3)X^{\gamma-1})\simeq0.\,
\end{equation}
Since $\dot\phi$ is now proportional to $\beta$, both the scalar
potential and the kinetic term are influenced, so typically
equations (\ref{motion7})-(\ref{motion8}) depend on $\beta$ even
though it is not explicitly present. The rest auxiliary parameters
and the slow-roll conditions as well, have the same forms as
before however they also appear to depend on $\beta$. The only
parameter which should be further amended, since it is of
paramount importance in this particular formalism, is the
$e$-folding number, which is essentially written in terms of the
constant-roll parameter as,
\begin{equation}
\centering
\label{efolds2}
N=\frac{1}{1-\beta}\int_{\phi_i}^{\phi_f}{\frac{\xi''}{\xi'}}\, ,
\end{equation}

\subsection{Revisiting the Error Function Model with the Constant-Roll Evolution}

Let us now proceed with an example. Suppose that the Gauss-Bonnet
coupling is once again an error function, for convenience since it
simplifies the ratio $\xi'/\xi''$. Let,
\begin{equation}
\centering
\label{xiB}
\xi(\phi)=\frac{2\Lambda_2}{\sqrt{\pi}}\int_{0}^{\delta_2\kappa\phi}{e^{-x^2}dx}\, ,
\end{equation}
Then as a result, the scalar potential must have the following
form in order for the continuity equation to be satisfied,
\begin{equation}
\centering
\label{VB}
V(\phi)=\frac{(2 \delta_2 ^2 \kappa  \phi) ^2 \left(\alpha  \left(\beta ^2+2 \beta -3\right)+12 \delta_2 ^2\right)}{-(1-\beta)^3 (\beta +1) \omega }\, ,
\end{equation}
which is simply a power-law potential with a constraint scalar
amplitude. Essentially, even if the Gauss-Bonnet scalar coupling
choice seems bizarre, the resulting scalar potential is elegant.
In consequence, the slow-roll conditions along with the auxiliary
parameters are now,
\begin{equation}
\centering
\label{index1B}
\epsilon_1=-\frac{(1-\beta)^2 \left((1-\beta)^2 \omega  V(\phi )-12 \alpha  \delta_2^4 \kappa ^2 \phi ^2\right)}{96 \delta_2^8 \kappa ^4 \phi ^4}\, ,
\end{equation}
\begin{equation}
\centering
\label{index5B}
\epsilon_5=\frac{4 (\beta -1) \kappa ^3 \Lambda_2 V(\phi )}{8 (\beta -1) \kappa ^3\Lambda_2 V(\phi )-3 \sqrt{\pi } \delta_2 \phi  e^{\delta_2^2 \kappa ^2 \phi ^2}}\, ,
\end{equation}
\begin{equation}
\centering
\label{index6B}
\epsilon_6=-\frac{2 (\beta -1)^2 \kappa  \Lambda_2 \left(-\phi  V'(\phi )+2 \delta_2^2 \kappa ^2 \phi ^2 V(\phi )+V(\phi )\right)}{\delta_2^2 \phi ^2 \left(8 (\beta -1) \kappa ^3 \Lambda_2 V(\phi )-3 \sqrt{\pi } \delta_2 \phi  e^{\delta_2^2 \kappa ^2 \phi ^2}\right)}\, ,
\end{equation}
\begin{equation}
\centering
\label{esb}
\epsilon_s=\frac{4 (1-\beta ) \kappa ^3 \Lambda_2 V(\phi ) e^{-\delta_2^2 \kappa ^2 \phi ^2}}{3 \sqrt{\pi }\delta_2 \phi }-\frac{3 \left(\frac{(1-\beta )^4 \omega  V(\phi )^2}{288 \delta_2^8 \kappa ^4 \phi ^4}-\frac{\alpha  (1-\beta )^2 V(\phi )}{24 \delta_2^4 \kappa ^2 \phi ^2}\right)}{V(\phi )}\, ,
\end{equation}
\begin{equation}
\centering
\label{deltaxiB}
\delta_\xi=\frac{(\beta -1) \kappa ^3 \Lambda_2 V(\phi ) e^{-\delta_2^2 \kappa ^2 \phi ^2}}{3 \sqrt{\pi } \delta_2 \phi }\, ,
\end{equation}
\begin{equation}
\centering
\label{lambdaB}
\lambda=-\frac{(\beta -1)^4 \omega  V(\phi )^2}{288 \delta_2^8 \kappa ^4 \phi ^4}\, ,
\end{equation}
Here we presented only a sample of the auxiliary parameters. From
the simple form of index $\epsilon_1$, given that the potential is
also a power-law, the final value of the scalar field of the
inflationary era can be derived. Furthermore, since the
Gauss-Bonnet coupling was chosen on purpose so as the ratio
$\xi'/\xi''$ is simplified, the initial value of the scalar field
during the first horizon crossing is also derivable. The values of
the scalar field for the case at hand are,
\begin{equation}
\centering
\label{phiiB}
\phi_i= \frac{\sqrt{\delta_2 ^2 \kappa ^2 \phi_f^2+(1-\beta)N}}{\delta_2  \kappa }\, ,
\end{equation}
\begin{equation}
\centering
\label{phifB}
\phi_f=\pm\frac{ \sqrt{1-\beta } \sqrt{-\alpha  \beta ^2+\alpha  \beta +6 \delta_2 ^2}}{\sqrt{12 \beta  \delta_2 ^4 \kappa ^2+12 \delta_2 ^4 \kappa ^2}}\, ,
\end{equation}
Designating ($\omega$, $\alpha$, $\Lambda_2$, $\delta_2$, $N$,
$\beta$)=(10, 1, 1, 0.4, 60, 0.001) produces compatible with the
observations results as $n_S=0.966941$, $r=0.0135517$,
$n_T=-0.0165296$ and $f_{NL}^{eq}=-22.8587$ are acceptable values.
The non-Gaussianities are obviously enhanced, since
$f_{NL}^{eq}=-22.8587$ in this case. Furthermore, the field
propagation velocity in the initial and final stage has the
numerical values of $c_s= 0.102481$ and $c_s=0.100141$
respectively, which suggests a decrease but more importantly
predicts no Jeans instabilities, hence the model is free of
ghosts. The behavior of the inflationary phenomenology parameters
and of the sound wave speed can be found in Figs. 3 and
4, for various sets of values of the free parameters.
The slow-roll indices obtain the values $\epsilon_1=0.0082648$,
$\epsilon_5=\epsilon_6=2\cdot10^{-27}$ which is practically zero
and moreover the auxiliary slow-varying parameters are $
\delta_\xi=-5\cdot10^{-28}$, $s=5\cdot10^{-18}$,
$\eta_s=0.0165289$ and $\lambda=-0.00645602$. It should be noted
here that even though $\beta\ll1$, the overall phenomenology is
quite different between the slow-roll and constant-roll cases,
something that can easily be ascertained from the numerical values
of the auxiliary parameters, for instance the field propagation
velocity. A notable feature of the model is that $f_{NL}^{eq}$ is
quite large, thus the primordial power spectrum of this model has
enhanced non-Gaussianities, and actually to a large magnitude.
This is also the main difference between the canonical and
$k$-inflation-corrected models.
\begin{figure}[h!]
\centering
\label{plot3}
\includegraphics[width=20pc]{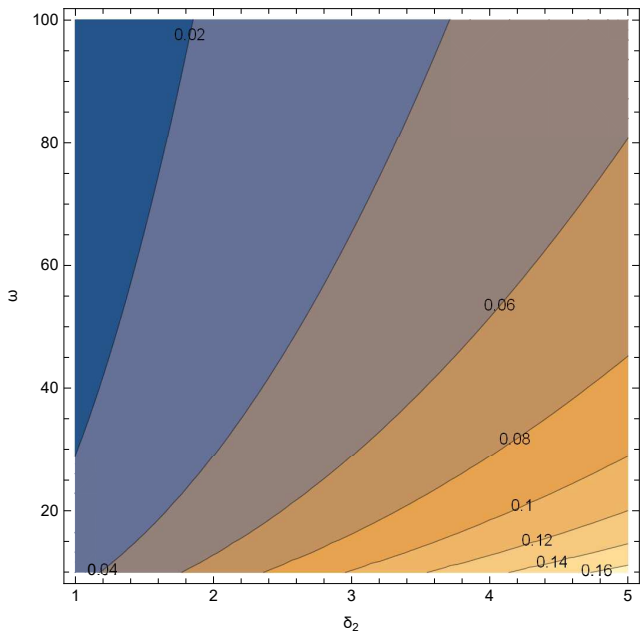}
\caption{Tensor-to-scalar ratio $r$ depending on the non canonical
parameter $\omega$ and exponent $\delta_2$ ranging from $[10,100]$
and [1,5] respectively. Apart from the acceptable values, it is
clear that there exists also pairs which produce incompatible with
the observations tensor-to-scalar ratio.}
\end{figure}

\begin{figure}[h!]
\centering
\label{plot4}
\includegraphics[width=20pc]{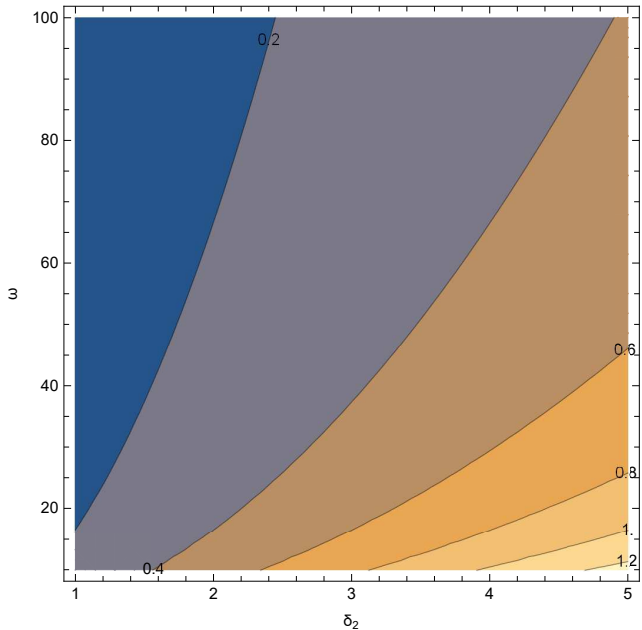}
\caption{Field propagation velocity $c_s$ during the first horizon
crossing depending on parameters $\omega$ and $\delta_2$ ranging
from $[10,100]$ and [1,5] respectively. For certain pairs, the
resulting velocity is greater than unity, meaning causality is
violated and hence are unacceptable. Moreover, these pairs also
produce an unacceptable value for the tensor-to-scalar ratio so
they are also excluded from an observation perspective.}
\end{figure}
When it comes to the parameter $\alpha$, we mention that during
the first horizon crossing, either values $\alpha=-1, 0$ are
capable of producing viable results but during the final stage,
the field velocity becomes complex, therefore with this approach,
$\alpha$ gives rise to ghosts. Exponent $\delta_2$ affects the
results, in particular increasing its value results in a decrease
to the tensor-to-scalar ratio and a subsequent increase to the
equilateral non-linear term, while the scalar spectral index
remains invariant. The constant-roll parameter on the other hand
affects only the scalar spectral index $n_S$ and the amount of
non-Gaussianities, meaning $f_{NL}^{eq}$. Essentially, increasing
$\beta$ leads to a subsequent increase in $n_S$ while it lowers
the value of $f_{NL}^{eq}$. Finally, $\omega$ influences only the
tensor-to-scalar ratio and the equilateral non-linear term.
Decreasing its order of magnitude leads to an enhanced
$f_{NL}^{eq}$ while it decreases $r$. Also, $\omega$ must be
negative in order to avoid the emergence of ghosts. Finally, for
the approximations made, we mention that $\dot
H/H^2\sim\mathcal{O}(10^{-3})$, $\alpha X+\omega
X^\gamma\sim\mathcal{O}(10^{-2})$ and $V\sim\mathcal{O}(1)$.
Furthermore, $24\dot\xi H^3\sim\mathcal{O}(10^{-26})$, $16\dot\xi
H\dot H\sim\mathcal{O}(10^{-28})$ and
$24\xi'H^4\sim\mathcal{O}(10^{-25})$ while
$V'\sim\mathcal{O}(10^{-1})$, hence all the approximations indeed
apply.

As in the slow-roll case, let us examine the impact of the free parameters
on the observed indices and the sound wave velocity. As shown in Figs. 3 and
4, $\delta_2$ and $\omega$ are once again the decisive factors on $r$ and $c_s$.
Furthermore, the Gauss-Bonnet dimensionless parameter, $\lambda_2$ in this case,
does not influence the results, apart from the sound wave velocity during the
ending stage on inflation and once again only by extreme values. It stands to
reason that the two model have similar dependencies on $\omega$ and $\delta_2$.
The new addition, the constant-roll parameter seems to be quite interesting. Here,
$\beta$is not prohibited to obtain large values, in fact $\beta=0.2$ for instance seems
to produce results which are in agreement with recent observations, as $n_S=0.9674$,
$r=0.01472$ and $f_{NL}=-18.2796$. Here, the amount of non Gaussianities is decreased,
and no violation of causality is observed neither during the first horizon crossing not the
end of the inflationary era. For larger values of the constant-roll parameter, the scalar spectral
index passes the threshold and no longer resides in the area of acceptance. This is also one of
the reasons a small value for the constant-roll parameter was chosen in the first place, so as
to achieve compatibility for $n_S$, enhance $f_{NL}^{eq}$ and obtain a slow-roll index
$\epsilon_2$ of order $\mathcal{O}(10^{-3})$, meaning a small value as the rest slow-roll
indices.  The opposite applies when the constant-roll parameter is negative and therefore,
$\beta=-0.3$ can enhance further the non Gaussianities and produce viable sound wave velocities
but the scalar spectral index is once again outside of the bound. Thus, compatibility can indeed
be achieved with the Planck 2018 collaboration once the constant-roll parameter is roughly
of order $\mathcal{O}(10^{-1})$ and below.

Similar to the previous case, the value of $\gamma=-1$ is more
than capable of producing compatible with the observations results
while simultaneously being free of ghosts. Also, between the
slow-roll and constant-roll evolution, there exists no significant
difference in the case on non-Gaussianities. Lastly, the reader
should not be discouraged and think that this formalism works only
under the error function Gauss-Bonnet coupling. In principle,
there exists no limitation however the error function is quite
convenient given that the ratio $\xi'/\xi''$ appears in our
calculations. Other couplings such as a power-law and an
exponential should also work.

\section{Conclusions}

Studying the dynamics of an Einstein-Gauss-Bonnet gravity
accompanied by a non-canonical kinetic term during the
inflationary era has revealed many interesting features. Firstly,
the realization that the gravitational waves propagate through
spacetime with the velocity of light, makes it abundantly clear
that the scalar functions of the theory, in this case the scalar
potential and the Gauss-Bonnet coupling, are interconnected, or in
other words cannot be freely designated. Furthermore, the kinetic
term is strongly dependent on the Gauss-Bonnet coupling, meaning
that the non-canonical kinetic part is itself interconnected with
string corrections. In this paper, we showed that the
$k$-inflation-corrected Einstein-Gauss-Bonnet gravity, under both
the slow-roll and constant-roll conditions, is capable of
producing viable phenomenology since the observational quantities
are compatible with the latest observations coming from the Planck
2018 data. Since a non-canonical kinetic term is present, the
amount of non-Gaussianities, as expected, is enhanced but still
resides in the area of acceptance. The only obstacle is this
framework is the field propagation velocity $c_s$ which can lead
to Jeans instabilities or superluminal velocities. As
demonstrated, the choice of a quadratic kinetic term is capable of
producing a ghost free model without violating causality, by
choosing appropriately the values of the free parameters.
Moreover, certain models may seem to lead to viable phenomenology.
Finally, we mention that the slow-roll approach for the scalar
field is not mandatory and in fact the constant-roll assumption is
also an option capable for producing viable results. Undoubtedly,
the same principles are expected to apply in the case of a
non-minimal coupling between the Ricci scalar and the scalar
field. Finally, an important outcome of this work is that the
primordial non-Gaussianities in the power spectrum, are enhanced
in the presence of the $k$-inflation corrections, for both the
constant and slow-roll conditions. The magnitude of the
equilateral momentum non-linear term $f_{NL}^{eq}$ in the
constant-roll case, is reportable and quite large in magnitude.

\end{document}